\documentclass[journal=nalefd,manuscript=article,reprint,
layout=twocolumn,amsmath,amssymb,]{achemso}
\usepackage{mathrsfs}
\usepackage{graphicx}
\graphicspath{{./images/}}
\usepackage{subfigure}  
\usepackage{dcolumn}
\usepackage{bm}
\usepackage{amsmath,mathrsfs}    
\usepackage{upgreek}
\usepackage{textcomp}
\usepackage[T1]{fontenc}
\usepackage{hyperref}   
\usepackage{comment}
\usepackage{siunitx}

\title{Scanning Gate Microscopy of Localized States in a gate-defined Bilayer Graphene Channel}
\author{Carolin Gold}
\email{cgold@phys.ethz.ch}
\author{Annika Kurzmann}
\affiliation{Solid State Physics Laboratory, ETH Zürich, CH-8093 Zürich, Switzerland}
\author{Kenji Watanabe}
\affiliation{Research Center for Functional Materials, 
National Institute for Materials Science, 1-1 Namiki, Tsukuba 305-0044, Japan}
\author{Takashi Taniguchi}
\affiliation{International Center for Materials Nanoarchitectonics, 
National Institute for Materials Science,  1-1 Namiki, Tsukuba 305-0044, Japan}
\author{Klaus Ensslin}
\author{Thomas Ihn}
\affiliation{Solid State Physics Laboratory, ETH Zürich, CH-8093 Zürich, Switzerland}

\date{\today}

\begin{document}

\begin{abstract}
We use Scanning Gate Microscopy to demonstrate the presence of localized states arising from potential inhomogeneities in a \SI{50}{nm}-wide, gate-defined conducting channel in encapsulated bilayer graphene. When imaging the channel conductance under the influence of a local tip-induced potential, we observe ellipses of enhanced conductance as a function of the tip position. These ellipses allow us to infer the location of the localized states and to study their dependence on the displacement field. For large displacement fields, we observe that localized states tend to occur halfway into the channel. All our observations can be well explained within the framework of stochastic Coulomb blockade. 
\end{abstract}

\maketitle

\section{Introduction}
\label{sec:intro}
Graphene is a promising material for semiconducting quantum dot-based qubits due to its small spin-orbit coupling and small hyperfine interaction \cite{TrauzettelSpinqubitsgraphene2007}.
For a long time, the fabrication of high-quality graphene-based nanodevices was hindered, e.g., by the presence of strong background potential modulations arising from the inhomogeneities and surface roughness of the silicon oxide substrates, as well as the existence of localized states at the etched device boundaries \cite{PonomarenkoChaoticDiracBilliard2008,StampferTunableGrapheneSingle2008,BischoffElectronictripledottransport2013,BischoffLocalizedchargecarriers2015}. Only recently, the conjunction of two key technological advancements has enabled the reproducible fabrication of high-quality graphene nanodevices. First, using encapsulation between hexagonal boron nitride (hBN) \cite{DeanBoronnitridesubstrates2010}, edge-contacting \cite{WangOneDimensionalElectricalContact2013} and graphite backgates \cite{OverwegElectrostaticallyInducedQuantum2018} enabled the fabrication of single- and bilayer graphene devices with very high electronic quality. Second, the use of bilayer graphene with its electrostatically tunable bandgap provided the possibility to create gate-defined devices \cite{OostingaGateinducedinsulatingstate2008}. These devices require an average disorder potential smaller than the gate-induced band-gap. If the latter condition is met, their smooth edge potentials in comparison to the previously etched devices minimizes the appearance of localized edge states. This made it possible to prepare high-quality quantum point contacts \cite{OverwegElectrostaticallyInducedQuantum2018} as well as few-electron or -hole quantum dots in bilayer-graphene channels \cite{EichSpinValleyStates2018}. These devices allow observing a plethora of intriguing phenomena such as a tunable valley splitting in quantum point contacts \cite{LeeTunableValleySplitting2020a}, excited states in few-electron or few-hole quantum dots \cite{KurzmannExcitedStatesBilayer2019}, and charge detection \cite{KurzmannChargeDetectionGateDefined2019}. However, the variability in the conductance properties of nominally identical devices fabricated in close proximity to each other indicates that local inhomogeneity of the electrostatic potential and disorder may still play an important role. To date, no experiments exist that focus on the local and systematic study of inhomogeneities and the possible formation of localized states in such devices.

In this paper we use Scanning Gate Microscopy (SGM) to demonstrate the formation of localized states in a gate-defined bilayer graphene channel. Scanning the voltage-biased metallic tip of an atomic force microscope over the channel perturbs the potential landscape locally. This allows us to probe the inhomogeneity of the system along the channel axis. To make the effects of inhomogeneities more pronounced, we investigate a channel which is narrower ($\SI{50}{nm}$) than the typical channel-widths used for quantum dot studies ($\SI{100}{nm}$). If the conductance through the channel is almost completely suppressed in the absence of the tip, we observe conductance resonances associated with localized states. These resonances lead to concentric rings or ellipses in the two-dimensional conductance maps obtained in SGM measurements. The center points of these ellipses allow us to infer the position of the localized states. By varying the displacement field forming the channel, we furthermore gain insights into the evolution of the localized states as a function of the band-gap size and electron concentration.

\section{Sample-characterization}
\label{sec: charact_no_tip}

\begin{figure*}
	\includegraphics[scale=1]{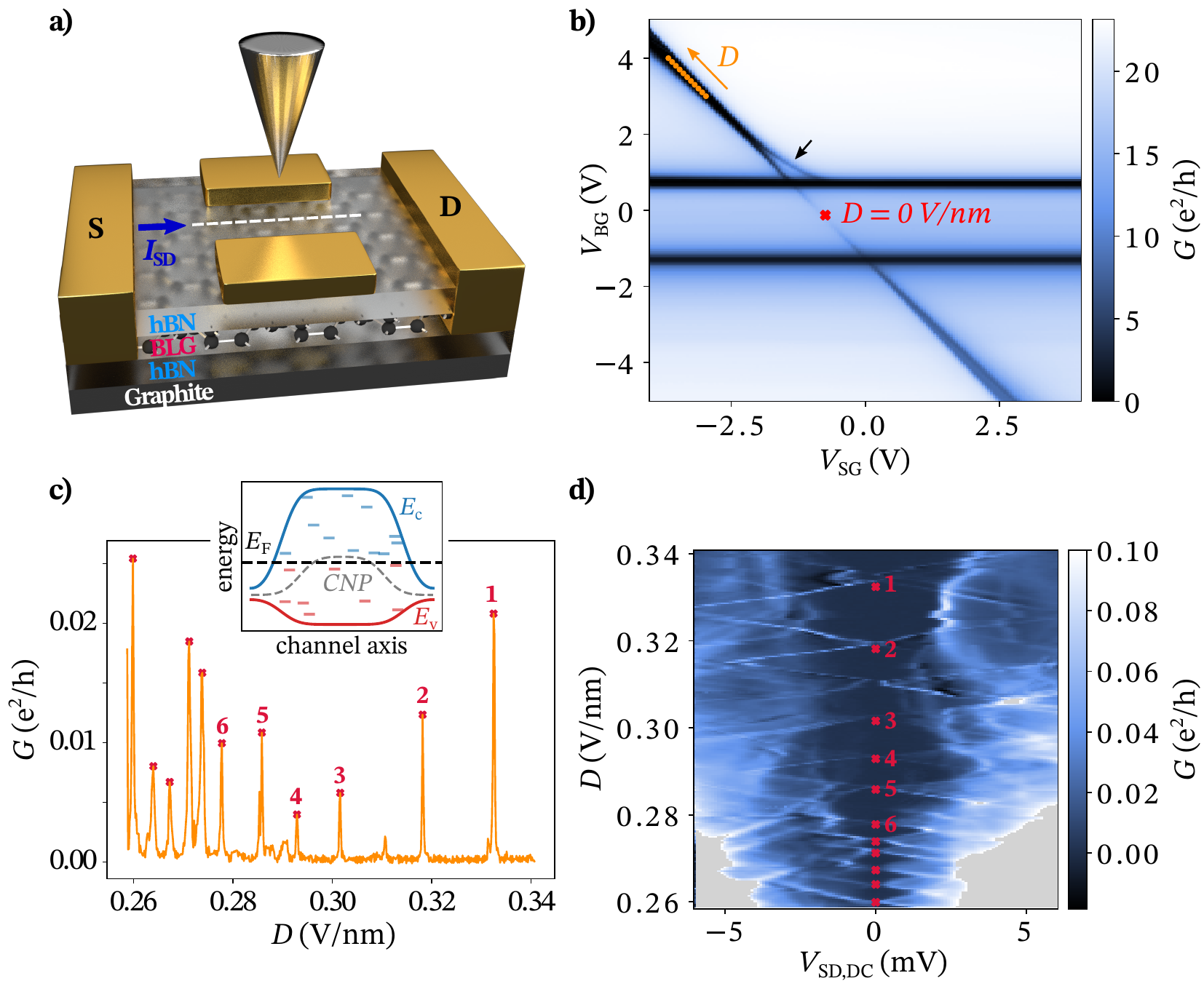}
	\caption{(a) Schematic of the measurement setup consisting of sample plus SGM-tip. The sample consists of an encapsulated bilayer graphene (BLG) on a graphite backgate. Displacement fields between split-gates and graphite backgate form a channel. The voltage biased AFM tip is scanned $\SI{55}{nm}$ (with respect to the top hBN-layer surface) above the channel while we measure the conductance $G$ between source (S) and drain (D). (b) Two-terminal linear conductance G as a function of split-gate voltage $V_\mathrm{SG}$ and back-gate voltage $V_\mathrm{BG}$ in absence of the tip. The diagonal line corresponds to charge neutrality underneath the split-gates. The displacement field in the double-gated areas increases along the dark-orange arrow with respect to $D=\SI{0}{V/nm}$ denoted by the red cross. (c) Conductance as a function of the displacement field $D$ along the charge neutrality point (orange dotted line in Fig.~\ref{fig: Fig1}b). The inset depicts the schematic of a possible potential landscape along the channel axis (cf dashed, white line in Fig.~\ref{fig: Fig1}a). $E_\mathrm{c}$ ($E_\mathrm{v}$) denotes the conduction (valence) band, $E_\mathrm{F}$ the Fermi-energy and $CNP$ the charge neutrality point.  (d) Coulomb diamonds.}
	\label{fig: Fig1}
\end{figure*}

We perform measurements on the van-der Waals stack depicted schematically in Fig. \ref{fig: Fig1}a. It consists of a bilayer graphene (BLG) flake encapsulated between two hBN flakes of different thicknesses $t$ ($t_\mathrm{top-hBN}=\SI{32}{nm}$, $t_\mathrm{bottom-hBN}=\SI{40}{nm}$) and deposited on top of a graphite backgate (BG). The fabrication of the device follows the process outlined in Ref. \citenum{OverwegElectrostaticallyInducedQuantum2018} . Ohmic source (S) and drain (D) contacts connect to the BLG-sheet while lithographically defined split-gates (SG, $\SI{300}{nm}$ long and $\SI{50}{nm}$ apart) allow us to confine the current flow in the graphene to a channel between the split-gates. Throughout this work, the same voltage $V_{\mathrm{SG}}$ is applied to both split-gates, and the back gate bias is given by the voltage $V_{\mathrm{BG}}$. A displacement field $D=[C_\mathrm{BG}(V_\mathrm{BG}-V_\mathrm{BG}^0)-C_\mathrm{SG}(V_\mathrm{SG}-V_\mathrm{SG}^0)]/2\epsilon_0$ is generated in the double-gated areas by applying a voltage difference between split- (SG) and backgate (BG). Here, $C_\mathrm{SG}$ and $C_\mathrm{BG}$ are the respective gate-graphene capacitances per unit area of the split-gate (SG) and back gate (BG). The charge neutrality point in the double-gated area (cf red cross in Fig. \ref{fig: Fig1}b) occurs at the offset voltages $V_\mathrm{SG}^0=\SI{-0.744}{V}$ and $V_\mathrm{BG}^0=\SI{-0.13}{V}$. As the gate voltages are tuned away from these voltages, the thus generated displacement field opens a bandgap in the bilayer graphene below the split-gated region and in the channel. The average of split- and backgate voltage weighted by the gate capacitances furthermore allows us to tune the Fermi level in the graphene sheet below the split-gate into the band gap, rendering it insulating. This results in a narrow channel consisting of the remaining conductive region between the gates.

All our measurements are performed at ${T=\SI{270}{mK}}$ in a $^3$He-cryostat. By applying both an AC voltage $V_\mathrm{SD,AC}=\SI{50}{\upmu V_\mathrm{rms}}$ as well as a variable DC voltage $V_\mathrm{SD,DC}$  between the source (S) and drain (D) contacts, we can perform both linear and nonlinear transport measurements. To this end, we measure the differential conductance $G=I_\mathrm{SD,AC}/V_\mathrm{SD,AC}$ using a home-built IV-converter and a low-frequency lock-in.
To explore the local properties of the channel, we perform SGM-measurements. As depicted in Fig.~\ref{fig: Fig1}a, we thus position a voltage-biased metallic tip at a height $h_\mathrm{tip}=\SI{55}{nm}$ above the top hBN surface in the area above the channel. By raster-scanning the tip we measure the differential conductance $G$ as a function of the tip position $\left(x,y\right)$ or the tip voltage $V_\mathrm{tip}$.

To demonstrate the formation of the channel experimentally, we first characterize the combined effect of the backgate voltage $V_\mathrm{BG}$ and split-gate voltage $V_\mathrm{SG}$ on the conductance through the device. Figure~\ref{fig: Fig1}b depicts the linear conductance $G(V_\mathrm{SG},V_\mathrm{BG})$. The diagonal conductance minimum depending on both $V_{\mathrm{BG}}$ and $V_{\mathrm{SG}}$ corresponds to charge neutrality of the area underneath the split-gates. Its depth depends non-monotonic on the displacement field $D$, reflecting the electrostatic landscape in the channel. The zero of the displacement field is the point of maximum conductance along this line, which is marked by the red cross in Fig.~\ref{fig: Fig1}b. For all gate-voltages along the conductance minimum, the split-gated areas have zero total density and tend to be increasingly insulating for increasing displacement fields due to the increasing band-gap in the dual-gated regions. Therefore the source-drain current is confined to the channel at sufficiently large displacement fields along this minimum.

In addition, we observe two horizontal, split-gate independent conductance minima. The small additional feature marked by the black arrow in Fig.~\ref{fig: Fig1}b indicates that the conductance minimum at positive $V_{\mathrm{BG}}$ most likely arises due to the charge neutrality point of the bulk regions around the channel. The minimum at negative $V_{\mathrm{BG}}$ belongs to the charge neutrality point of additional devices connected in series to that of interest here, which are not used in the present work.

In contrast to previous measurements on samples with mostly larger channel widths \cite{OverwegElectrostaticallyInducedQuantum2018, EichElectrostaticallyDefinedQuantum2019}, the formation of the channel not only reduces the conductance but suppresses it almost completely for the displacement fields marked by the orange dotted line in Fig. \ref{fig: Fig1}b. Along this line, the bulk regions of the charge carrier gas outside the channel are n-type with a density larger than \SI{1e12}{cm^{-2}}, while the double-gated areas are at zero density with the Fermi level in the middle of the band gap. 
In contrast to the double-gated regions, the channel region only couples to the split-gates via stray fields. As the electrostatic coupling of the channel is thus dominated by the back-gate, the electron density in the channel increases with increasing displacement field along the orange line of reduced conductance (cf $D$-arrow in Fig.~\ref{fig: Fig1}b).
The experimental fact that the device reaches zero conductance at positive displacement field values suggests that charge neutrality in the channel occurs at these positive displacement fields. This is further corroborated by the asymmetric suppression of the conductance along the diagonal line with respect to positive and negative displacement fields $D$.

Measuring the conductance along the orange dotted line in Fig.~\ref{fig: Fig1}b with high resolution reveals the conductance resonances of different spacing and height shown in Fig.~\ref{fig: Fig1}c. To identify the same peaks throughout the measurements presented in this paper, we label them from one to six. The observation of such strong conductance resonances suggests that localized states exist inside the channel region, which transmit resonantly once their energy coincides with the Fermi level.

This interpretation is confirmed by the finite bias spectroscopy measurements in Fig.~\ref{fig: Fig1}d, which shows diamonds of suppressed conductance and irregular size. They represent the fingerprints of one or more localized states in the channel. 
The presence of low-energy excited state resonances outside the diamonds of suppressed conductance supports the interpretation in terms of Coulomb blockade. The diamond between resonances 1 and 2 has an extent of \SI{6.3}{mV} in $V_{\mathrm{SD,DC}}$-direction. Assuming that the entire voltage drops across this single quantum dot-like impurity leads to an energy calibration factor for the displacement field axis of $\SI{420}{meV/(V/nm)}$. Figure~\ref{fig: Fig1}d therefore spans an energy range of about \SI{34}{meV}, which is the same order of magnitude as typical energy gaps expected at these displacement fields \cite{McCannElectronicPropertiesMonolayer2012,OverwegElectrostaticallyinducednanostructures2018,RickhausTransportNetworkTopological2018}.

The charging energy for increasing electron numbers grows from $\SI{2}{meV}$ to more than $\SI{6}{meV}$ with increasing displacement field. This finding suggests that the channel tends to become more insulating with increasing displacement field, presumably due to the increasing energy gap and the Fermi energy moving closer to the center of the gap.

The schematic in the inset of Fig.~\ref{fig: Fig1}c summarizes the present understanding of the device along the white dashed line in Fig.~\ref{fig: Fig1}a. 
Here, $E_\mathrm{v}$ ($E_\mathrm{c}$) denotes the energy of the valence (conduction) band edge along the channel axis, $E_\mathrm{F}$ is the Fermi-energy and $CNP$ the charge neutrality point. Inhomogeneities of the background potential induce localized states in the gap as indicated schematically by the horizontal lines within the band gap. Depending on their position with respect to the charge neutrality point, these localized states can have stronger valence band character (light-red lines in the inset of Fig~\ref{fig: Fig1}c) or stronger conduction band character (light-blue lines in the inset of Fig.~\ref{fig: Fig1}c). Electrons occupying these states may tunnel from the channel into the high-density regions outside the channel.
Localized states half-way along the channel axis will typically have the highest transmission in the case where a single state dominates the conductance of the whole channel.

In such a model, localized states would be distributed in an uncontrolled way both in position and energy throughout the channel.
The model supports the idea of stochastic Coulomb blockade\cite{KemerinkStochasticCoulombblockade1994}, where one or more localized states participate in transport through the channel, depending on their positions and the mismatch between their respective quantized energy levels.

\section{Tip-influence on the conductance resonances}
\label{sec:charact_with_tip}

\begin{figure*}
	\includegraphics[scale=1]{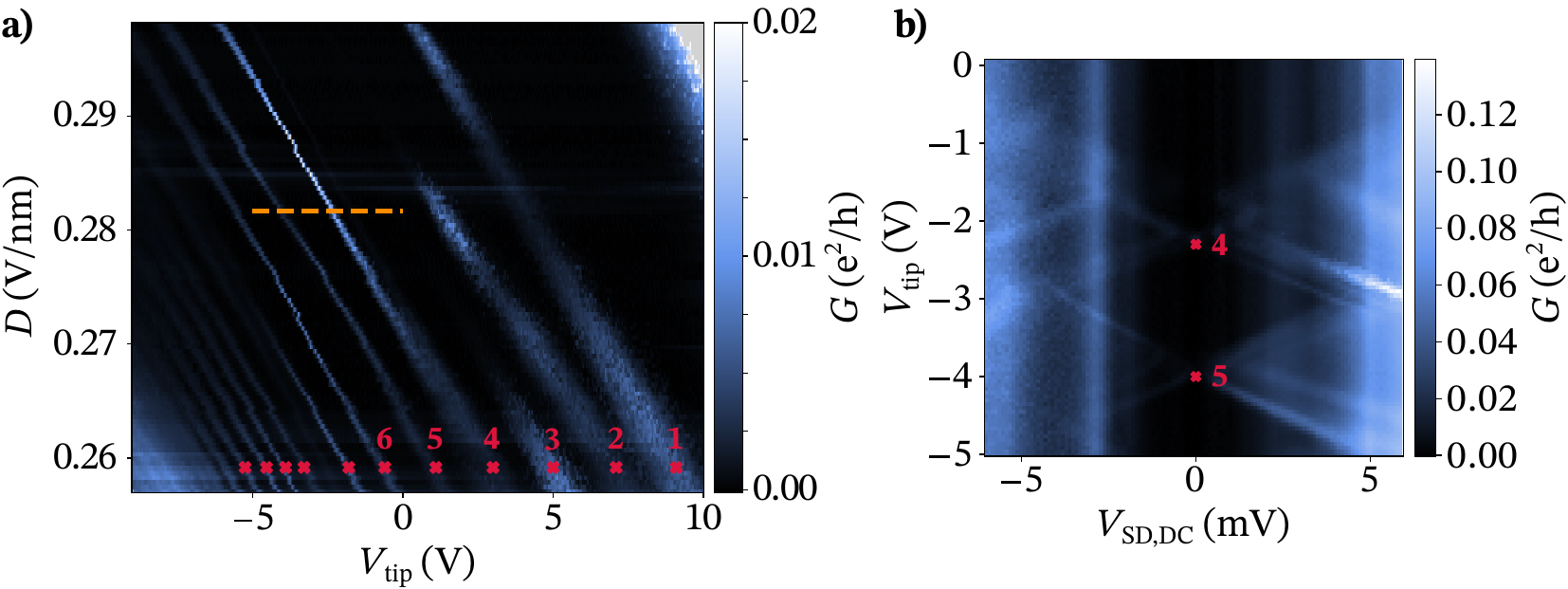}
	\caption{Measurement at fixed lateral tip position above the channel. (a) two-terminal conductance as a function of tip voltage $V_\mathrm{tip}$ and displacement field $D$ along the charge neutrality point (cf orange line in Fig.\ref{fig: Fig1}b). (b) Coulomb diamonds measured as a function of tip voltage $V_\mathrm{tip}$ at the displacement field denoted by the orange dashed line in Fig. \ref{fig: Fig2}a.}
	\label{fig: Fig2}
\end{figure*}

\begin{figure*}
	\includegraphics[scale=1]{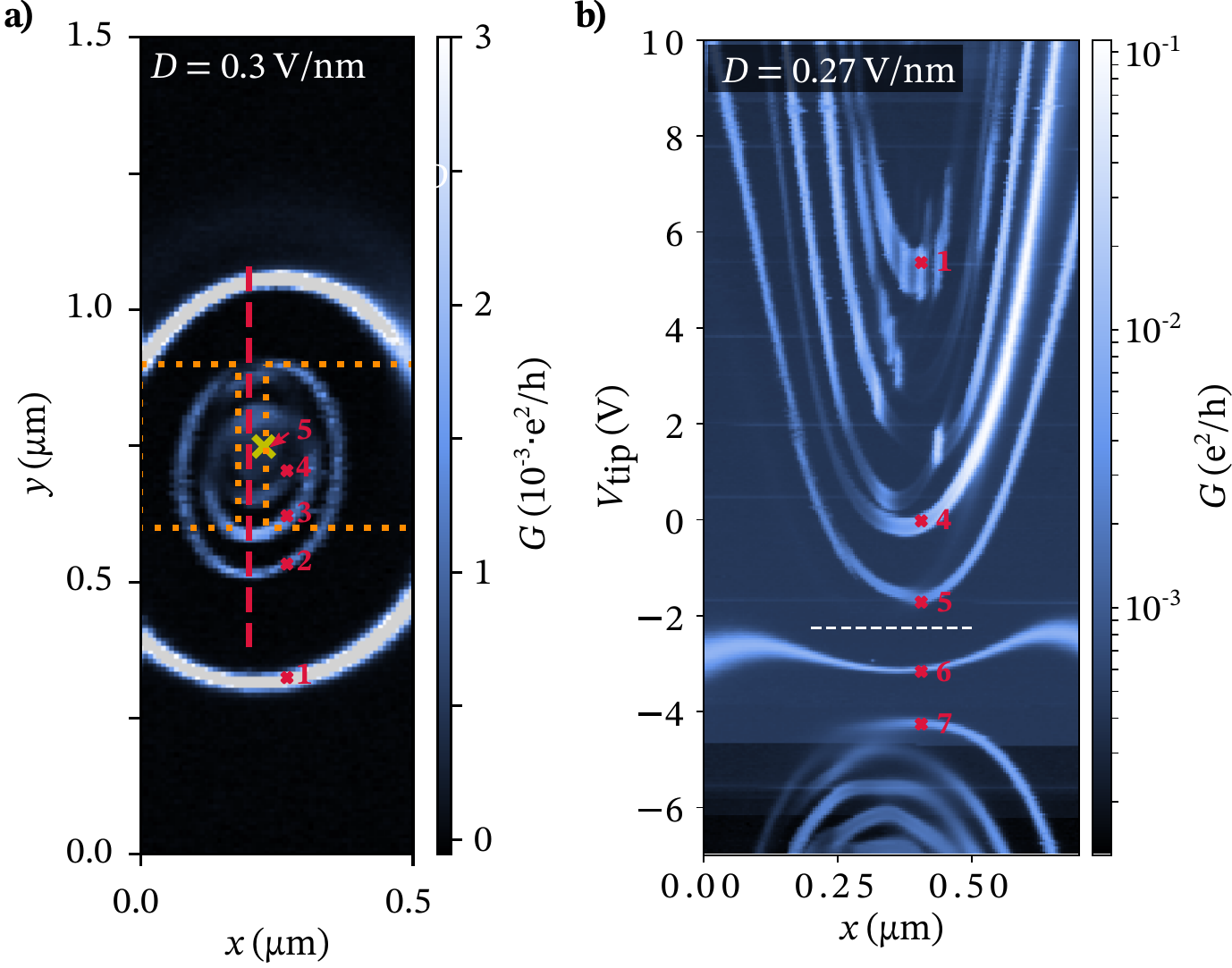}
	\caption{(a) SGM measurement ($V_\mathrm{tip}=\SI{-8}{V}$, $D=\SI{0.3}{V/nm}$) above the channel depicting concentric rings around the middle of the channel. The outline of the split-gates are denoted by the orange dashed lines. The yellow cross marks the tip position at which the  data in Fig. \ref{fig: Fig2} is obtained. (b) Logarithmic plot of the linear conductance G(V$_\mathrm{tip}$,x) along the red dashed line in Fig. \ref{fig: Fig2}a for a displacement field $D=\SI{0.27}{V/nm}$ different from that in Fig. \ref{fig: Fig3}a. The white dashed line denotes the approximate position of the split-gates.}
	\label{fig: Fig3}
\end{figure*}

To investigate the formation of the localized states in the channel, we perform SGM measurements, initially at a fixed lateral tip position above the channel (cf yellow cross in Fig. \ref{fig: Fig3}a). We first measure the linear conductance $G$ as a function of tip-voltage $V_\mathrm{tip}$ and displacement field $D$. The resulting data depicted in Fig. \ref{fig: Fig2}a show that the conductance resonances depend linearly on both tip voltage and displacement field as expected from a capacitance argument. The resonances labeled in Fig.~\ref{fig: Fig1}c can be identified using the strong correlation of the peak positions along a vertical cut at $V_\mathrm{tip}=\SI{-5}{V}$, which we call the least-invasive tip voltage, with the data in Fig.~\ref{fig: Fig1}c. The resonances in Fig.~\ref{fig: Fig2}a are labeled accordingly.
We attribute the offset of the least-invasive tip-voltage from zero to the difference in the work-functions between the PtIr-tip and the sample.

The negative slope of the resonances in Fig.~\ref{fig: Fig2}a together with the fact that increasing tip voltage will increase the number of electrons in the channel confirms that increasing the displacement field increases the electron number in the channel.
Moving along the orange dotted line in Fig.~\ref{fig: Fig1}b in positive $D$-direction, the split gate exactly compensates the electron-accumulating tendency of the back gate in dual-gated regions of the device. In the channel, however, the action of the split gate is reduced to stray-field effects, and as a consequence, electron accumulation by the back gate gains dominance.

The observation that all resonances in Fig.~\ref{fig: Fig2}a have essentially the same slope indicates that the resonances originate either from one localized state or from few spatially closely spaced localized states.
The increasing separation of resonances for increasingly positive tip voltage is in agreement with the same observation in Fig.~\ref{fig: Fig1}d.

Individual resonances would die out in Fig.~\ref{fig: Fig2}a with increasing $D$ and decreasing $V_{\mathrm{tip}}$, because the increase in $D$ also acts in reducing the tunnel coupling of the states to the leads, while $V_{\mathrm{tip}}$ locally keeps the localized states resonant with the Fermi level.

The influence of the tip at fixed position on the energy of the localized states allows us to measure Coulomb blockade diamonds as a function of tip-voltage and dc-source drain bias as depicted in Fig. \ref{fig: Fig2}b. Similar to Fig. \ref{fig: Fig1}d we observe Coulomb blockade diamonds of different sizes, which are  consistent with a constant lever arm and the increasing zero-bias peak spacing. From these measurements we extract a tip lever arm ${\alpha_{\mathrm{tip}}=0.0023}$. This rather small lever arm is the result of the screening action of the split gate.

In order to obtain spatial information about the position of the localized states in the channel, we raster scan the SGM tip in the region above the channel. For these measurements, the channel is formed by the displacement field $D=\SI{0.3}{V/nm}$ and the tip voltage is given by $V_{\mathrm{tip}}=\SI{-8}{V}$, which is repulsive for electrons. The resulting measurement is shown in Fig.~\ref{fig: Fig3}a. We observe  concentric rings centered roughly around the location of the localized states.

The ring-shape of the resonances can be understood as follows: Due to the extended electrostatic potential of the tip induced in the channel, a change in the distances between the voltage biased tip and the localized states has a similar effect on the energy levels as a change in the tip voltage. The absolute separation between the tip and the localized states therefore allows us to tune their energy levels with respect to the Fermi energy, and thereby to observe conductance resonances. The shape of a specific resonant ring is dictated by the geometry of the potential induced by the tip in the graphene plane, which is itself determined by the tip and gate geometries. The spherical geometry of the tip together with the extended long edges of the split-gate leads to a nearly elliptic shape of individual resonant rings, with the long axis of the ellipses along the channel axis.

For rings at large lateral tip--localized state separations the position of the localized states cannot be determined with great accuracy. However, bringing the tip closer to the center of the channel
we observe two rings centered at slightly different positions close to the center of the channel. This observation confirms that resonant transport occurs through states localized in the channel mid-way along the channel axis.

To ascertain the peak-number belonging to these ellipses, we extract the peak occurring at $D\approx\SI{0.3}{V/nm}$ and $V_\mathrm{tip}=\SI{-8}{V}$ in Fig.~\ref{fig: Fig2}a as peak number 5. As the measurement depicted in Fig.~\ref{fig: Fig2}a was obtained at the tip position marked by the yellow cross in Fig.~\ref{fig: Fig3}a, the innermost ellipse close to this tip position thus corresponds to peak 5. Due to the decrease of the tip-potential influence for increasing tip-localized state distances, this allows us to identify the remaining ellipses from the inner ellipse outwards as labeled in Fig. \ref{fig: Fig3}a. 

To get a better understanding of the influence of the tip potential and the position of the localized states that dominate transport, we measure the conductance $G\left(x,V_\mathrm{tip}\right)$ along the channel axis (red dashed line in Fig. \ref{fig: Fig3}a). The gating effect of the tip gives rise to the parabolic resonances observed in Fig. \ref{fig: Fig3}b. The curvature of those parabolas changes from convex to concave at the symmetry line $V_\mathrm{tip}\approx\SI{-3}{V}$, which we call the least-invasive tip voltage. It is offset from $V_\mathrm{tip}^{LI}=\SI{-5}{V}$ observed in Fig.~\ref{fig: Fig2}a due to small charge-rearrangements in the sample over time. The change from convex to concave in the resonance at $V_\mathrm{tip}=\SI{-3}{V}$ occurs exactly at the outer corners of the channel gates, outlined by the white-dashed line.  Following each parabola individually, we observe that their extrema occur at different tip positions $x$ along the dashed red line. This is in good agreement with the observation that the concentric rings in Fig. \ref{fig: Fig3}a are centered at different positions in the center of the channel. 
Resonances at tip voltages below the least-invasive voltage $V_\mathrm{tip}^{LI}=\SI{-3}{V}$ unambiguously show avoided crossings indicating mutual interactions between the respective localized states. This observation further strengthens the notion of stochastic Coulomb blockade.

\begin{figure*}
	\includegraphics[scale=1]{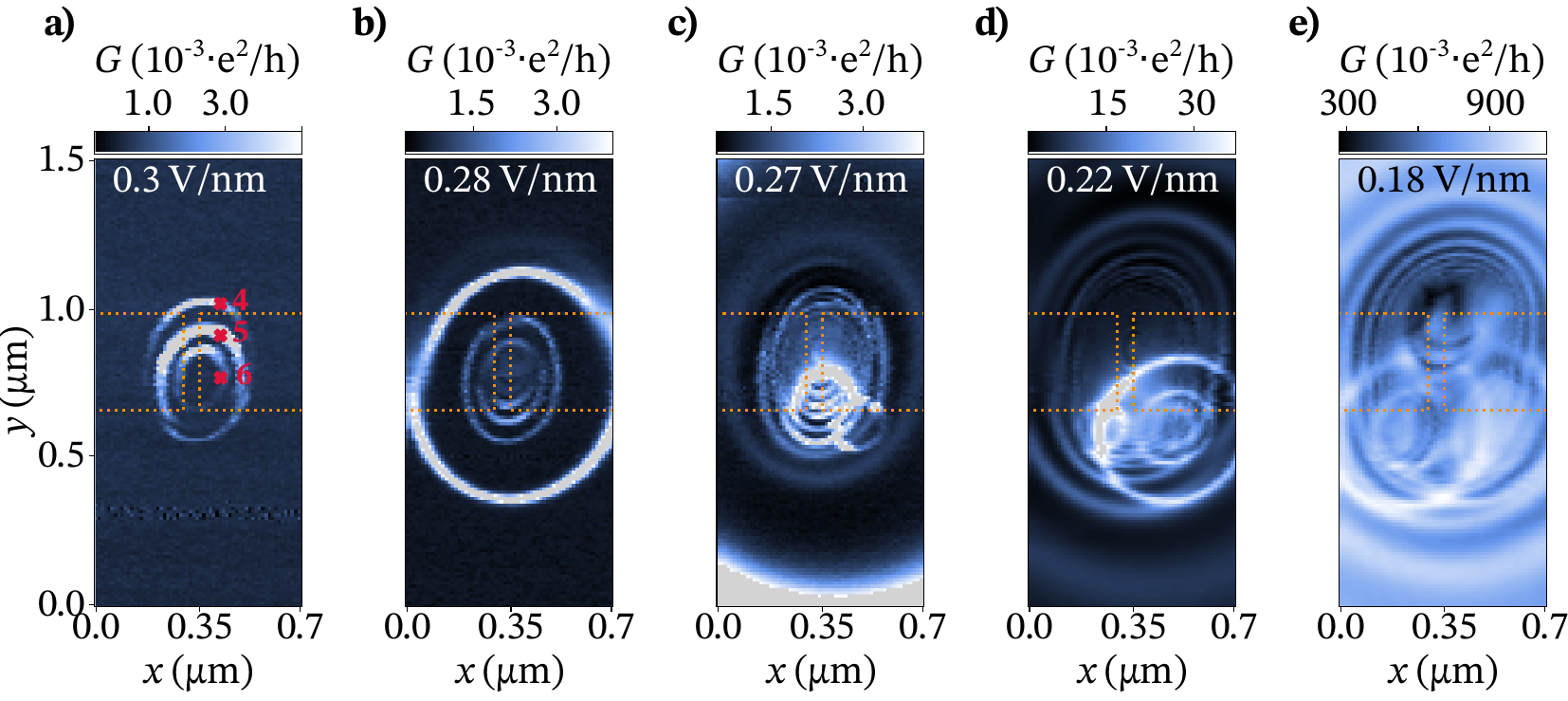}
	\caption{Evolution of the SGM measurements (V$_\mathrm{tip}=\SI{-9}{V}$) as a function of displacement field $D$ (upper white text-inset). The orange dashed lines denote the outline of the split-gates. }
	\label{fig: Fig4}
\end{figure*}


In order to investigate the localized states observed in Fig. \ref{fig: Fig2}a for specific stepwise decreased displacement fields, we perform two-dimensional scans for $V_\mathrm{tip}=\SI{-9}{V}$. The results are depicted in Fig. \ref{fig: Fig4}. The leftmost figure (Fig.~\ref{fig: Fig4}a) depicts the measurement for $D=\SI{0.3}{V/nm}$ which shows three concentric ellipses centered at approximately the same position close to the middle of the channel. Identifying the peaks with the method used in Fig.~\ref{fig: Fig3}a, the inner ellipse corresponds to peak number 6. This allows us to label the middle and outer ring with peak number 5 respectively 4. The observation that ellipses are centered in one single location could correspond to a single localized state in the center of the channel.

Decreasing the displacement field to $D=\SI{0.28}{V/nm}$ (Fig.~\ref{fig: Fig4}b), we observe (at least) two additional elliptical resonances. The smaller ellipses close to the center of the channel are centered at two different positions within the channel, whereas the larger outer ellipses seem to be centered at the same position within the channel. A reliable identification of the peak numbers in this configuration ($D=\SI{0.28}{V/nm}$, $V_\mathrm{tip}=\SI{-9}{V}$) is hindered by the very dense peak-spacing in this configuration in Fig.~\ref{fig: Fig2}a.
However, similar to the SGM measurement in Fig.~\ref{fig: Fig3}a, the concentric rings are preferentially centered at positions close to the middle of the channel rather than in the outer channel regions.

Decreasing the displacement field even further, more and more concentric ellipses appear in the SGM measurements (cf Fig.~\ref{fig: Fig4}c-e) . 
In addition to the ellipses centered in the middle of the channel, we observe rings centered at additional positions close to the edges of the split-gates for displacement fields $D\leq\SI{0.27}{V/nm}$ (cf Fig.~\ref{fig: Fig4}c-e). The first of those rings appears at $D=\SI{0.27}{V/nm}$ (Fig.~\ref{fig: Fig4}c) and their number increases for even smaller displacement fields ($D=\SI{0.22}{V/nm}$ in Fig.~\ref{fig: Fig4}d and $D=\SI{0.18}{V/nm}$ in Fig.~\ref{fig: Fig4}e). We interpret their appearance as evidence that the conductance below the split gates starts to increase at these displacement fields. This increase appears to be spatially inhomogeneous, dominated by  individual localized states.

\section{Conclusion}
We presented scanning gate measurements on a gate-defined channel in high quality bilayer graphene. The scanning gate images reveal the occurrence of stochastic Coulomb blockade in the conductance through the channel at various displacement fields. We have shown that the conjunction of the split-gates forming the channel and the back gate is sufficient to render the $\SI{50}{nm}$-wide channel insulating. Our data suggests that the charge neutrality point of the channel in our sample is located at positive displacement fields. This suggests that transport in the region below $D=\SI{0.34}{V/nm}$ occurs through states of a character dominated by the valence band. In the situation of strongly suppressed conductance, localized states are most likely to be observed close to the middle of the channel. If the displacement field becomes too small, the dual-gated regions start to leak. This leakage is again dominated by local inhomogeneity of the potential.

\section{Acknowledgments}
We thank Peter Märki as well as the staff of the ETH cleanroom facility FIRST for their technical support. We also acknowledge financial support by the ETH Zurich grant ETH-38 17-2 and the European Graphene Flagship. Growth of hexagonal boron nitride crystals was supported by the Elemental Strategy Initiative conducted by the MEXT, Japan, Grant Number JPMXP0112101001, JSPS KAKENHI Grant Number JP20H00354 and the CREST(JPMJCR15F3), JST.

\bibliography{Library_P04_Channel_QD_sub}

\providecommand{\latin}[1]{#1}
\makeatletter
\providecommand{\doi}
  {\begingroup\let\do\@makeother\dospecials
  \catcode`\{=1 \catcode`\}=2 \doi@aux}
\providecommand{\doi@aux}[1]{\endgroup\texttt{#1}}
\makeatother
\providecommand*\mcitethebibliography{\thebibliography}
\csname @ifundefined\endcsname{endmcitethebibliography}
  {\let\endmcitethebibliography\endthebibliography}{}
\begin{mcitethebibliography}{19}
\providecommand*\natexlab[1]{#1}
\providecommand*\mciteSetBstSublistMode[1]{}
\providecommand*\mciteSetBstMaxWidthForm[2]{}
\providecommand*\mciteBstWouldAddEndPuncttrue
  {\def\EndOfBibitem{\unskip.}}
\providecommand*\mciteBstWouldAddEndPunctfalse
  {\let\EndOfBibitem\relax}
\providecommand*\mciteSetBstMidEndSepPunct[3]{}
\providecommand*\mciteSetBstSublistLabelBeginEnd[3]{}
\providecommand*\EndOfBibitem{}
\mciteSetBstSublistMode{f}
\mciteSetBstMaxWidthForm{subitem}{(\alph{mcitesubitemcount})}
\mciteSetBstSublistLabelBeginEnd
  {\mcitemaxwidthsubitemform\space}
  {\relax}
  {\relax}

\bibitem[Trauzettel \latin{et~al.}(2007)Trauzettel, Bulaev, Loss, and
  Burkard]{TrauzettelSpinqubitsgraphene2007}
Trauzettel,~B.; Bulaev,~D.~V.; Loss,~D.; Burkard,~G. Spin Qubits in Graphene
  Quantum Dots. \emph{Nature Physics} \textbf{2007}, \emph{3}, 192--196\relax
\mciteBstWouldAddEndPuncttrue
\mciteSetBstMidEndSepPunct{\mcitedefaultmidpunct}
{\mcitedefaultendpunct}{\mcitedefaultseppunct}\relax
\EndOfBibitem
\bibitem[Ponomarenko \latin{et~al.}(2008)Ponomarenko, Schedin, Katsnelson,
  Yang, Hill, Novoselov, and Geim]{PonomarenkoChaoticDiracBilliard2008}
Ponomarenko,~L.~A.; Schedin,~F.; Katsnelson,~M.~I.; Yang,~R.; Hill,~E.~W.;
  Novoselov,~K.~S.; Geim,~A.~K. Chaotic {{Dirac Billiard}} in {{Graphene
  Quantum Dots}}. \emph{Science} \textbf{2008}, \emph{320}, 356--358\relax
\mciteBstWouldAddEndPuncttrue
\mciteSetBstMidEndSepPunct{\mcitedefaultmidpunct}
{\mcitedefaultendpunct}{\mcitedefaultseppunct}\relax
\EndOfBibitem
\bibitem[Stampfer \latin{et~al.}(2008)Stampfer, Schurtenberger, Molitor,
  G{\"u}ttinger, Ihn, and Ensslin]{StampferTunableGrapheneSingle2008}
Stampfer,~C.; Schurtenberger,~E.; Molitor,~F.; G{\"u}ttinger,~J.; Ihn,~T.;
  Ensslin,~K. Tunable {{Graphene Single Electron Transistor}}. \emph{Nano
  Letters} \textbf{2008}, \emph{8}, 2378--2383\relax
\mciteBstWouldAddEndPuncttrue
\mciteSetBstMidEndSepPunct{\mcitedefaultmidpunct}
{\mcitedefaultendpunct}{\mcitedefaultseppunct}\relax
\EndOfBibitem
\bibitem[Bischoff \latin{et~al.}(2013)Bischoff, Varlet, Simonet, Ihn, and
  Ensslin]{BischoffElectronictripledottransport2013}
Bischoff,~D.; Varlet,~A.; Simonet,~P.; Ihn,~T.; Ensslin,~K. Electronic
  Triple-Dot Transport through a Bilayer Graphene Island with Ultrasmall
  Constrictions. \emph{New Journal of Physics} \textbf{2013}, \emph{15},
  083029\relax
\mciteBstWouldAddEndPuncttrue
\mciteSetBstMidEndSepPunct{\mcitedefaultmidpunct}
{\mcitedefaultendpunct}{\mcitedefaultseppunct}\relax
\EndOfBibitem
\bibitem[Bischoff \latin{et~al.}(2015)Bischoff, Varlet, Simonet, Eich, Overweg,
  Ihn, and Ensslin]{BischoffLocalizedchargecarriers2015}
Bischoff,~D.; Varlet,~A.; Simonet,~P.; Eich,~M.; Overweg,~H.~C.; Ihn,~T.;
  Ensslin,~K. Localized Charge Carriers in Graphene Nanodevices. \emph{Applied
  Physics Reviews} \textbf{2015}, \emph{2}, 031301\relax
\mciteBstWouldAddEndPuncttrue
\mciteSetBstMidEndSepPunct{\mcitedefaultmidpunct}
{\mcitedefaultendpunct}{\mcitedefaultseppunct}\relax
\EndOfBibitem
\bibitem[Dean \latin{et~al.}(2010)Dean, Young, Meric, Lee, Wang, Sorgenfrei,
  Watanabe, Taniguchi, Kim, Shepard, and Hone]{DeanBoronnitridesubstrates2010}
Dean,~C.~R.; Young,~A.~F.; Meric,~I.; Lee,~C.; Wang,~L.; Sorgenfrei,~S.;
  Watanabe,~K.; Taniguchi,~T.; Kim,~P.; Shepard,~K.~L.; Hone,~J. Boron Nitride
  Substrates for High-Quality Graphene Electronics. \emph{Nature
  Nanotechnology} \textbf{2010}, \emph{5}, 722--726\relax
\mciteBstWouldAddEndPuncttrue
\mciteSetBstMidEndSepPunct{\mcitedefaultmidpunct}
{\mcitedefaultendpunct}{\mcitedefaultseppunct}\relax
\EndOfBibitem
\bibitem[Wang \latin{et~al.}(2013)Wang, Meric, Huang, Gao, Gao, Tran,
  Taniguchi, Watanabe, Campos, Muller, Guo, Kim, Hone, Shepard, and
  Dean]{WangOneDimensionalElectricalContact2013}
Wang,~L.; Meric,~I.; Huang,~P.~Y.; Gao,~Q.; Gao,~Y.; Tran,~H.; Taniguchi,~T.;
  Watanabe,~K.; Campos,~L.~M.; Muller,~D.~A.; Guo,~J.; Kim,~P.; Hone,~J.;
  Shepard,~K.~L.; Dean,~C.~R. One-{{Dimensional Electrical Contact}} to a
  {{Two}}-{{Dimensional Material}}. \emph{Science} \textbf{2013}, \emph{342},
  614--617\relax
\mciteBstWouldAddEndPuncttrue
\mciteSetBstMidEndSepPunct{\mcitedefaultmidpunct}
{\mcitedefaultendpunct}{\mcitedefaultseppunct}\relax
\EndOfBibitem
\bibitem[Overweg \latin{et~al.}(2018)Overweg, Eggimann, Chen, Slizovskiy, Eich,
  Pisoni, Lee, Rickhaus, Watanabe, Taniguchi, Fal'ko, Ihn, and
  Ensslin]{OverwegElectrostaticallyInducedQuantum2018}
Overweg,~H.; Eggimann,~H.; Chen,~X.; Slizovskiy,~S.; Eich,~M.; Pisoni,~R.;
  Lee,~Y.; Rickhaus,~P.; Watanabe,~K.; Taniguchi,~T.; Fal'ko,~V.; Ihn,~T.;
  Ensslin,~K. Electrostatically {{Induced Quantum Point Contacts}} in {{Bilayer
  Graphene}}. \emph{Nano Letters} \textbf{2018}, \emph{18}, 553--559\relax
\mciteBstWouldAddEndPuncttrue
\mciteSetBstMidEndSepPunct{\mcitedefaultmidpunct}
{\mcitedefaultendpunct}{\mcitedefaultseppunct}\relax
\EndOfBibitem
\bibitem[Oostinga \latin{et~al.}(2008)Oostinga, Heersche, Liu, Morpurgo, and
  Vandersypen]{OostingaGateinducedinsulatingstate2008}
Oostinga,~J.~B.; Heersche,~H.~B.; Liu,~X.; Morpurgo,~A.~F.; Vandersypen,~L.
  M.~K. Gate-Induced Insulating State in Bilayer Graphene Devices. \emph{Nature
  Materials} \textbf{2008}, \emph{7}, 151--157\relax
\mciteBstWouldAddEndPuncttrue
\mciteSetBstMidEndSepPunct{\mcitedefaultmidpunct}
{\mcitedefaultendpunct}{\mcitedefaultseppunct}\relax
\EndOfBibitem
\bibitem[Eich \latin{et~al.}(2018)Eich, Pisoni, Overweg, Kurzmann, Lee,
  Rickhaus, Ihn, Ensslin, Herman, Sigrist, Watanabe, and
  Taniguchi]{EichSpinValleyStates2018}
Eich,~M.; Pisoni,~R.; Overweg,~H.; Kurzmann,~A.; Lee,~Y.; Rickhaus,~P.;
  Ihn,~T.; Ensslin,~K.; Herman,~F.; Sigrist,~M.; Watanabe,~K.; Taniguchi,~T.
  Spin and {{Valley States}} in {{Gate}}-{{Defined Bilayer Graphene Quantum
  Dots}}. \emph{Physical Review X} \textbf{2018}, \emph{8}, 031023\relax
\mciteBstWouldAddEndPuncttrue
\mciteSetBstMidEndSepPunct{\mcitedefaultmidpunct}
{\mcitedefaultendpunct}{\mcitedefaultseppunct}\relax
\EndOfBibitem
\bibitem[Lee \latin{et~al.}(2020)Lee, Knothe, Overweg, Eich, Gold, Kurzmann,
  Klasovika, Taniguchi, Wantanabe, Fal'ko, Ihn, Ensslin, and
  Rickhaus]{LeeTunableValleySplitting2020a}
Lee,~Y.; Knothe,~A.; Overweg,~H.; Eich,~M.; Gold,~C.; Kurzmann,~A.;
  Klasovika,~V.; Taniguchi,~T.; Wantanabe,~K.; Fal'ko,~V.; Ihn,~T.;
  Ensslin,~K.; Rickhaus,~P. Tunable {{Valley Splitting}} Due to {{Topological
  Orbital Magnetic Moment}} in {{Bilayer Graphene Quantum Point Contacts}}.
  \emph{Physical Review Letters} \textbf{2020}, \emph{124}, 126802\relax
\mciteBstWouldAddEndPuncttrue
\mciteSetBstMidEndSepPunct{\mcitedefaultmidpunct}
{\mcitedefaultendpunct}{\mcitedefaultseppunct}\relax
\EndOfBibitem
\bibitem[Kurzmann \latin{et~al.}(2019)Kurzmann, Eich, Overweg, Mangold, Herman,
  Rickhaus, Pisoni, Lee, Garreis, Tong, Watanabe, Taniguchi, Ensslin, and
  Ihn]{KurzmannExcitedStatesBilayer2019}
Kurzmann,~A.; Eich,~M.; Overweg,~H.; Mangold,~M.; Herman,~F.; Rickhaus,~P.;
  Pisoni,~R.; Lee,~Y.; Garreis,~R.; Tong,~C.; Watanabe,~K.; Taniguchi,~T.;
  Ensslin,~K.; Ihn,~T. Excited {{States}} in {{Bilayer Graphene Quantum Dots}}.
  \emph{Physical Review Letters} \textbf{2019}, \emph{123}, 026803\relax
\mciteBstWouldAddEndPuncttrue
\mciteSetBstMidEndSepPunct{\mcitedefaultmidpunct}
{\mcitedefaultendpunct}{\mcitedefaultseppunct}\relax
\EndOfBibitem
\bibitem[Kurzmann \latin{et~al.}(2019)Kurzmann, Overweg, Eich, Pally, Rickhaus,
  Pisoni, Lee, Watanabe, Taniguchi, Ihn, and
  Ensslin]{KurzmannChargeDetectionGateDefined2019}
Kurzmann,~A.; Overweg,~H.; Eich,~M.; Pally,~A.; Rickhaus,~P.; Pisoni,~R.;
  Lee,~Y.; Watanabe,~K.; Taniguchi,~T.; Ihn,~T.; Ensslin,~K. Charge
  {{Detection}} in {{Gate}}-{{Defined Bilayer Graphene Quantum Dots}}.
  \emph{Nano Letters} \textbf{2019}, \emph{19}, 5216--5221\relax
\mciteBstWouldAddEndPuncttrue
\mciteSetBstMidEndSepPunct{\mcitedefaultmidpunct}
{\mcitedefaultendpunct}{\mcitedefaultseppunct}\relax
\EndOfBibitem
\bibitem[Eich(2019)]{EichElectrostaticallyDefinedQuantum2019}
Eich,~M. Electrostatically {{Defined Quantum Dots}} in {{Bilayer Graphene}}.
  Doctoral {{Thesis}}, ETH Zurich, 2019\relax
\mciteBstWouldAddEndPuncttrue
\mciteSetBstMidEndSepPunct{\mcitedefaultmidpunct}
{\mcitedefaultendpunct}{\mcitedefaultseppunct}\relax
\EndOfBibitem
\bibitem[McCann(2012)]{McCannElectronicPropertiesMonolayer2012}
McCann,~E. In \emph{Graphene {{Nanoelectronics}}: {{Metrology}}, {{Synthesis}},
  {{Properties}} and {{Applications}}}; Raza,~H., Ed.; {{NanoScience}} and
  {{Technology}}; {Springer}: {Berlin, Heidelberg}, 2012; pp 237--275\relax
\mciteBstWouldAddEndPuncttrue
\mciteSetBstMidEndSepPunct{\mcitedefaultmidpunct}
{\mcitedefaultendpunct}{\mcitedefaultseppunct}\relax
\EndOfBibitem
\bibitem[Overweg(2018)]{OverwegElectrostaticallyinducednanostructures2018}
Overweg,~H. Electrostatically Induced Nanostructures in Bilayer Graphene.
  Doctoral {{Thesis}}, ETH Zurich, 2018\relax
\mciteBstWouldAddEndPuncttrue
\mciteSetBstMidEndSepPunct{\mcitedefaultmidpunct}
{\mcitedefaultendpunct}{\mcitedefaultseppunct}\relax
\EndOfBibitem
\bibitem[Rickhaus \latin{et~al.}(2018)Rickhaus, Wallbank, Slizovskiy, Pisoni,
  Overweg, Lee, Eich, Liu, Watanabe, Taniguchi, Ihn, and
  Ensslin]{RickhausTransportNetworkTopological2018}
Rickhaus,~P.; Wallbank,~J.; Slizovskiy,~S.; Pisoni,~R.; Overweg,~H.; Lee,~Y.;
  Eich,~M.; Liu,~M.-H.; Watanabe,~K.; Taniguchi,~T.; Ihn,~T.; Ensslin,~K.
  Transport {{Through}} a {{Network}} of {{Topological Channels}} in {{Twisted
  Bilayer Graphene}}. \emph{Nano Letters} \textbf{2018}, \emph{18},
  6725--6730\relax
\mciteBstWouldAddEndPuncttrue
\mciteSetBstMidEndSepPunct{\mcitedefaultmidpunct}
{\mcitedefaultendpunct}{\mcitedefaultseppunct}\relax
\EndOfBibitem
\bibitem[Kemerink and Molenkamp(1994)Kemerink, and
  Molenkamp]{KemerinkStochasticCoulombblockade1994}
Kemerink,~M.; Molenkamp,~L.~W. Stochastic {{Coulomb}} Blockade in a Double
  Quantum Dot. \emph{Applied Physics Letters} \textbf{1994}, \emph{65},
  1012--1014\relax
\mciteBstWouldAddEndPuncttrue
\mciteSetBstMidEndSepPunct{\mcitedefaultmidpunct}
{\mcitedefaultendpunct}{\mcitedefaultseppunct}\relax
\EndOfBibitem
\end{mcitethebibliography}

\end{document}